\documentstyle[12pt]{article}

\begin{document}
\title{\bf {Vacuum Quantum Effects for Parallel Plates Moving by
Uniform Acceleration in Static de Sitter Space }}
\author{M.R. Setare  \footnote{E-mail: rezakord@ipm.ir}
  \\{Physics Dept. Inst. for Studies in Theo. Physics and
Mathematics(IPM)}\\
{P. O. Box 19395-5531, Tehran, IRAN }\\
 }
\date{\small{}}
 \maketitle

\begin{abstract}
The Casimir forces on two parallel plates moving by uniform proper
acceleration in static de Sitter background due to conformally
coupled massless scalar field satisfying Dirichlet boundary
conditions on the plates is investigated. Static de Sitter space
is conformally related to the Rindler space, as a result we can
obtain vacuum expectation values of energy-momentum tensor for
conformally invariant field in static de Sitter space from the
corresponding Rindler counterpart by the conformal transformation.

 \end{abstract}
\newpage

 \section{Introduction}

  The Casimir effect is one of the most interesting manifestations
  of nontrivial properties of the vacuum state in quantum field
  theory [1,2]. Since its first prediction by
  Casimir in 1948\cite{Casimir} this effect has been investigated for
  different fields having different boundary geometries[3-9]. The
  Casimir effect can be viewed as the polarization of
  vacuum by boundary conditions or geometry. Therefore, vacuum
  polarization induced by a gravitational field is also considered as
  Casimir effect.\\
    Casimir stress for parallel
   plates in the background of static domain wall in four and two dimensions
  is calculated in \cite{{set1},{set2}}. Spherical bubbles immersed in different de Sitter
  spaces in- and out-side is calculated in \cite{{set3},{set4}}.  \\
  It is well known that the vacuum
state for an uniformly accelerated observer, the Fulling--Rindler
vacuum \cite{Full73,Full77,Unru76,Boul75, sahrin}, turns out to be
inequivalent to that for an inertial observer, the familiar
Minkowski vacuum. Quantum field theory in accelerated systems
contains many of special features produced by a gravitational
field avoiding some of the difficulties entailed by
renormalization in a curved spacetime. In particular, near the
canonical horizon in the gravitational field, a static spacetime
may be regarded as a Rindler--like spacetime. Rindler space is
conformally related to the static de Sitter space and to the
Robertson--Walker space with negative spatial curvature. As a
result the expectation values of the energy--momentum tensor for a
conformally invariant field and for corresponding conformally
transformed boundaries on the de Sitter and Robertson--Walker
backgrounds can be derived from the corresponding Rindler
counterpart by the standard transformation  \cite{davies}. The
authors in \cite{davies} have shown that the Minkowski vacuum
contains a thermal spectrum of Rindler particles. One can also
demonstrate this by showing that the Green functions in Minkowski
vacuum are Rindler thermal Green functions. In a similar way one
can relate the vacua of static de Sitter space and de Sitter space
have the same curvature, but static de Sitter space is a member of
Rindler class, while de Sitter space is a member of Minkowski
space.\\
  In this paper we will study the scalar vacuum
polarization brought about by the presence of infinite plane
boundary moving by uniform acceleration through the static de
Sitter vacuum. This problem for the conformally coupled Dirichlet
and Neumann massless scalar and electromagnetic fields in four
dimensional Rindler spacetime was considered by Candelas and
Deutsch \cite{CandD}. Here we will investigate the vacuum
expectation values of the energy-momentum tensor for the massless
scalar field with conformal curvature coupling and satisfying
Dirichlet boundary condition on the infinite plane in four
spacetime dimension.
 Here we use
the results of Ref. \cite{sahrin} to generate vacuum
energy--momentum tensor for the static de Sitter background which
is conformally related to the Rindler spacetime. Previously this
method has been used in \cite{set5} to drive the vacuum stress on
parallel plates for scalar field with Dirichlet boundary condition
in de Sitter space. Also this method has been used in \cite{set6}
to derive the vacuum characteristics of the Casimir configuration
on background of conformally flat brane-world geometries for
massless scalar field
with Robin boundary condition on the plates.\\
   In section two we calculate the stress on
  two parallel plates with Dirichlet boundary conditions.
   The last section conclude and  summarize the results.

\section{Vacuum expectation values for the energy-momentum tensor}
We will consider a conformally coupled massless scalar field $%
\varphi (x)$ satisfying the following equation
\begin{equation}
\left( \nabla _{\mu }\nabla ^{\mu }+\frac{1}{6} R\right) \varphi
(x)=0, \label{fieldeq}
\end{equation}
on the background of a static form of de Sitter space-time. In Eq.
(\ref{fieldeq}) $\nabla _{\mu }$ is the operator of the covariant
derivative, and $R$ is the Ricci scalar for the de Sitter space.
\begin{equation}
R=\frac{12}{\alpha^{2}}.  \label{Riccisc}
\end{equation}
The static form of de Sitter space time which is conformally
related to the Rindler space time is given by  \cite{davies}
\begin{equation}\label{eqds}
ds^{2}=[1-(\frac{r^{2}}{\alpha^{2}})]dt^{2}-[1-(\frac{r^{2}}{\alpha^{2}})]^{-1}dr^{2}-r^{2}
(d\theta^{2}+\sin^{2} \theta d\phi^{2}),
\end{equation}
under the change of variable
$\frac{r}{\alpha}=r'(1+r'^{2})^{-1/2}$ we obtain
\begin{equation}\label{eqrds}
ds^{2}=(\alpha^{2}-r^{2})[\frac{dt^{2}}{\alpha^{2}}-\gamma
dr'^{2}-r'^{2} (d\theta^{2}+\sin^{2} \theta d\phi^{2})],
\end{equation}
with $\gamma=(1+r'^{2})^{-1}$, which is manifestly conformal to
following Rindler space time if $\eta=\frac{t}{\alpha}$
\begin{equation}\label{eqri}
ds^{2}=\zeta^{2}[d\eta^{2}-\gamma dr^{2}-r^{2}
(d\theta^{2}+\sin^{2} \theta
d\phi^{2})]=\xi^{2}d\tau^{2}-d\xi^{2}-dy^{2}-dz^{2},
\end{equation}
This metric can be written in coordinates $(\tau ,r, \theta ,\phi
) $ related to the previous ones by transformation
\begin{equation}\label{coordtrans}
\xi =\frac{R}{1-R\cos \theta },\quad y=\frac{R\sin \theta \cos
\phi }{1-R\cos \theta },\quad z=\frac{R\sin \theta \sin \phi
}{1-R\cos \theta },
\end{equation}
where
\begin{equation}\label{Rr}
R=R(r)=\frac{r}{\sqrt{1+r^2}} \ .
\end{equation}
In these coordinates the metric takes the form
\begin{equation}\label{Rind2}
ds^2_{{\mathrm{Rind}}}=\xi ^2 \left[ d\tau
^2-\frac{dr^2}{1+r^2}-r^2\left( d\theta ^2+\sin ^2 \theta d\phi
^2\right) \right] \ , \quad \xi =\xi (r,\theta ) \ .
\end{equation}
Let us denote the energy--momentum tensor in coordinates $(\tau
,\xi , y, z)$ as $T_{\mu \nu }$, and the same tensor in
coordinates $(\tau ,r, \theta ,\phi ) $ as $T'_{\mu \nu }$. The
tensor $T_{\mu \nu }$ has the structure
\begin{equation}\label{Tmunu}
T_{\mu }^{\nu }={\mathrm{diag}}(\varepsilon ,-p,-p_{\perp
},p_{\perp }) \ .
\end{equation}
By making use the standard transformation law for the second rank
tensors, for the nonzero components in the new coordinate system
one finds
\begin{eqnarray}\label{Tmunuprime}
T'_{rr}&=& R^{\prime 2}(r)\left( \frac{\partial \xi }{\partial
R}\right) ^2 \left( T_{\xi \xi }+\sin ^2\theta T_{yy}\right) \\
T'_{r\theta }&=& \frac{\partial \xi }{\partial R} R'(r)\left[
\frac{\partial \xi }{\partial \theta }\left( T_{\xi \xi }+\sin
^2\theta T_{yy}\right) +\xi \sin \theta \cos \theta T_{yy} \right]
\\
T'_{\theta \theta }&=& \left( \frac{\partial \xi }{\partial \theta
}\right) ^2T_{\xi \xi }^2+\left[ \frac{\partial }{\partial
\theta}(\xi \sin \theta )\right] ^2 T_{yy} \\
T'_{\phi \phi }&=& \xi ^2 \sin ^2\theta T_{yy}.
\end{eqnarray}
Our main interest in the present paper is to investigate the
vacuum
expectation values (VEV's) of the energy--momentum tensor for the field $%
\varphi (x)$ in the background of the above de Sitter space time
induced by two parallel plates moving with uniform proper
acceleration.we will consider the case of a scalar field
satisfying Dirichlet boundary condition on the surface of the
plates:
\begin{equation}
\varphi \mid _{\xi =\xi _{1}}=\varphi \mid _{\xi =\xi _{2}}=0
\label{Dboundcond}
\end{equation}
The presence of boundaries modifies the spectrum of the
zero--point fluctuations compared to the case without boundaries.
This results in the shift in the VEV's of the physical quantities,
such as vacuum energy density and stresses. This is the well known
Casimir effect. It can be shown that for a conformally coupled
scalar by using field equation (\ref{fieldeq}) the expression for
the energy--momentum tensor can be presented in the form
\begin{equation}
T_{\mu \nu }=\nabla _{\mu }\varphi \nabla _{\nu }\varphi -\frac{1}{6} \left[ \frac{%
g_{\mu \nu }}{2}\nabla _{\rho }\nabla ^{\rho }+\nabla _{\mu
}\nabla _{\nu }+R_{\mu \nu }\right] \varphi ^{2},  \label{EMT1}
\end{equation}
where $R_{\mu \nu }$ is the Ricci tensor. The quantization of a
scalar filed on background of metric Eq.(3) is standard. Let
$\{\varphi _{\alpha }(x),\varphi _{\alpha }^{\ast }(x)\}$ be a
complete set of orthonormalized positive and negative frequency
solutions to the field equation (\ref {fieldeq}), obying boundary
condition (\ref{Dboundcond}). By expanding the field operator over
these eigenfunctions, using the standard commutation rules and the
definition of the vacuum state for the vacuum expectation values
of the energy-momentum tensor one obtains
\begin{equation}
\langle 0|T_{\mu \nu }(x)|0\rangle =\sum_{\alpha }T_{\mu \nu }\{\varphi {%
_{\alpha },\varphi _{\alpha }^{\ast }\}},  \label{emtvev1}
\end{equation}
where $|0\rangle $ is the amplitude for the corresponding vacuum
state, and the bilinear form $T_{\mu \nu }\{{\varphi ,\psi \}}$ on
the right is determined by the classical energy-momentum tensor
(\ref{EMT1}). Instead of evaluating Eq. (\ref{emtvev1}) directly
on background of the curved metric, the vacuum expectation values
can be obtained from the corresponding Rindler space time results
for a scalar field $\bar{\varphi}$ by using the conformal
properties of the problem under consideration. Under the
conformal transformation $g_{\mu \nu }=\Omega ^{2}\bar{g}_{\mu \nu }$ the $%
\bar{\varphi}$ field will change by the rule
\begin{equation}
\varphi (x)=\Omega ^{-1}\bar{\varphi}(x),  \label{phicontr}
\end{equation}
where for metric Eq.(3) the conformal factor is given by $\Omega
=\frac{\sqrt{\alpha^{2}-r^{2}}}{\xi}$. The Casimir effect with
boundary conditions (\ref{Dboundcond}) on two parallel plates
moving with uniform proper acceleration on background of the
Rindler spacetime is investigated in Ref. \cite{sahrin} for a
scalar field with a Dirichlet and Neumann boundary condition. In
the case of a conformally coupled scalar the corresponding
regularized VEV's for the energy-momentum tensor in the region
between the plates have the form
\begin{equation}
\langle 0_D| T_{i}^{k}|0_D\rangle = A_3 \delta
_{i}^{k}\int_{0}^{\infty }dkk^{3}\int_{0}^{\infty }d\omega
\,\left\{ \frac{\sinh \pi \omega }{\pi }\,f^{(i)}[\tilde{D}
_{i\omega }(k\xi ,k\xi _{2})] - \frac{I_{\omega }(k\xi
_{1})}{I_{\omega }(k\xi _{2})}\frac{F^{(i)}[D_{\omega }(k\xi ,k\xi _{2})]}{%
D_{\omega }(k\xi _{1},k\xi _{2})}\right\} , \label{EMTDdiag1}
\end{equation}
where $|0_D\rangle $ is the amplitude for the Dirichlet vacuum
between the plates, and
\begin{equation}\label{Adnot}
  A_3=\frac{1}{2\pi ^{2}}.
\end{equation}
also we have introduced the notation
\begin{equation}
\tilde{D}_{i\omega }(k\xi ,k\xi _{2})=K_{i\omega }(k\xi )-\frac{%
K_{i\omega }(k\xi _{2})}{I_{i\omega }(k\xi _{2})}I_{i\omega }(k\xi
), \label{ztilda}
\end{equation}
and the functions $F^{(i)}[G(z)]$, $i=0,1,2,3$ are as following
\begin{equation}
F^{(i)}[G(z)]=f^{(i)}[G(z),\omega \rightarrow i\omega ].
\label{Ffunc}
\end{equation}
here for a given function $G(z)$ we use the notations
\begin{eqnarray}
f^{(0)}[G(z)] &=& \frac{1}{6} \left| \frac{dG(z)}{dz}%
\right| ^{2}+\frac{1 }{6z}\frac{d}{dz}|G(z)|^{2}+\frac{1}{6}\left[
1 +5\frac{\omega ^{2}}{z^{2}} \right]
|G(z)|^{2},  \label{f0} \\
f^{(1)}[G(z)] &=&-\frac{1}{2}\left| \frac{dG(z)}{dz}\right|
^{2}-\frac{1
}{6z}\frac{d}{dz}|G(z)|^{2}+\frac{1}{2}\left( 1-\frac{\omega ^{2}}{z^{2}}%
\right) |G(z)|^{2},  \label{f1} \\
f^{(i)}[G(z)] &=&-\frac{|G(z)|^{2}}{2}+ \frac{1}{6}
\left[ \left| \frac{dG(z)}{dz}\right| ^{2}+\left( 1-\frac{\omega ^{2}}{z^{2}}%
\right) |G(z)|^{2}\right] ;\quad i=2,3,  \label{f23}
\end{eqnarray}
where $G(z)=D_{i\omega }(z,k\xi _{2})$, which given by  following
expression,  and the indices 0,1 correspond to the coordinates
$\tau $, $\xi $ respectively,
\begin{equation}
D_{i\omega }(k\xi ,k\xi _{2})=I_{i\omega }(k\xi _{2})K_{i\omega
}(k\xi )-K_{i\omega }(k\xi _{2})I_{i\omega }(k\xi ).
\label{Deigfunc}
\end{equation}
Now let us present the VEV's (\ref{EMTDdiag1}) in the form
\begin{equation}\label{tmiuniu}
\langle 0| T_{i}^{k}| 0\rangle _D=\langle 0_{R}| T_{i}^{k}|
0_{R}\rangle +\langle T_{i}^{k}\rangle ^{(1b)}_D(\xi _{1},\xi
)+\langle T_{i}^{k}\rangle ^{(1b)}_D(\xi _{2},\xi )+\Delta \langle
T_{i}^{k}\rangle _D(\xi _{1},\xi _{2},\xi ),\quad \xi _{1}<\xi
<\xi _{2},
\end{equation}
where
\begin{equation}
\langle 0_{R}| T_{i}^{k}| 0_{R}\rangle =\frac{A_3\delta
_{i}^{k}}{\pi }\int_{0}^{\infty }dkk^{3}\int_{0}^{\infty }d\omega
\sinh \pi \omega \,f^{(i)}[K_{i\omega }(k\xi )]  \label{DFR}
\end{equation}
are the corresponding VEV's for the Fulling--Rindler vacuum
without boundaries, all divergences are contained in this part.
 These divergences can be regularized subtracting the
corresponding VEV's for the Minkowskian vacuum. The subtracted
VEV's
\begin{equation}\label{subRind}
  \langle T_{i}^{k}\rangle _{{\mathrm{sub}}}^{(R)}=\langle 0_R|T_{i}^{k}
  |0_R\rangle -\langle 0_M|T_{i}^{k}|0_M\rangle
\end{equation}
are investigated in a large number of papers,
\cite{sahrin,CandD,Saharian1,CandRaine,Davi77,Cand78,Troo79,Brow85,Brow86,%
Hill,Frol87,Dowk87,Pare93,More96}. For a massless scalar VEV's
(\ref{subRind}) can be presented in the form
\begin{equation}\label{subRindm0}
  \langle T_{i}^{k}\rangle _{{\mathrm{sub}}}^{(R)}=-\frac{\delta _i^k
  \xi ^{-d-1}}{2^{d-1}\pi ^{d/2}\Gamma (d/2)}\int _{0}^{\infty }
  \frac{\omega ^d g^{(i)}(\omega )d\omega }{e^{2\pi \omega }+(-1)^d}
\end{equation}
(the expressions for the functions $g^{(i)}(\omega )$ are given in
Ref. \cite{Saharian1}) correspond to the absence from the vacuum
of thermal distribution with standard temperature
$T=(2\pi\xi)^{-1}$. As we see from Eq. (\ref{subRindm0}), the
corresponding spectrum has non- Planckian form: the density of
states factor is not proportional to $\omega ^{d-1}d\omega $ where
$d$ is dimension of space . The spectrum takes the Planckian form
for conformally coupled scalars in $d=1,2,3$ with $g^{(0)}(\omega
)=-d g^{(i)}(\omega )=1$, $i=1,2,\ldots d$. In four dimensional
space time $d=3$ we have
\begin{equation}\label{subRindm01}
  \langle T_{i}^{k}\rangle _{{\mathrm{sub}}}^{(R)}=\frac{-1}{480 \pi^{2}\xi^{4}}
diag(1,-1/3,-1/3,-1/3).
\end{equation}
The boundary part term $\langle T_{i}^{k}\rangle _D^{(1b)}(\xi
_{1},\xi )$ are give by
\begin{equation}
\langle T_{i}^{k}\rangle _D^{(1b)}(\xi _{1},\xi )=-A_3\delta
_{i}^{k}\int_{0}^{\infty }dkk^{3}\int_{0}^{\infty }d\omega
\frac{I_{\omega }(k\xi _{1})}{K_{\omega }(k\xi
_{1})}F^{(i)}[K_{\omega }(k\xi )]  \label{D1platebound}
\end{equation}
is induced in the region $\xi >\xi _1$ by the presence of a single
plane boundary located at $\xi =\xi _{1}$. The expressions for the
boundary part $\langle T_{i}^{ k}\rangle _D^{(1b)}(\xi _{2},\xi )$
in the region $\xi <\xi _{2}$ are obtained from formulae
(\ref{D1platebound}) by replacing
\begin{equation} I_{\omega
}\rightarrow K_{\omega },\quad K_{\omega }\rightarrow I_{\omega
},\quad \xi _{1}\rightarrow \xi _{2},\quad \xi _{2}\rightarrow \xi
_{1}. \label{replace}
\end{equation}
Also
\begin{equation} \label{intterm1}
\Delta \langle T_{i}^{k}\rangle _D=-A_3\delta _i^k\int_0^\infty dk
k^3\int_0^\infty d\omega
I_\omega (k\xi _1)\left[ \frac{F^{(i)}[D_\omega (k\xi ,k\xi _2)]}{%
I_\omega (k\xi _2)D_\omega (k\xi _1,k\xi
_2)}-\frac{F^{(i)}[K_\omega (k\xi )]}{K_\omega (k\xi _1)}\right]
\end{equation}
is the 'interference' term.\\
 The vacuum energy-momentum tensor on static de Sitter space (sd) Eq.(3) is
obtained by the standard transformation law between conformally
related problems (see, for instance, \cite{davies}) and has the
form
\begin{equation}
\langle T_{\nu }^{\mu }\left[ g_{\alpha \beta }\right] \rangle _{{\rm sd}%
}=\xi^{4}( \alpha^{2}-r^{2})^{-2}\langle T_{\nu }^{'\mu }\left[
g_{\alpha \beta }\right] \rangle _{{\rm Rindler}}+\langle T_{\nu
}^{\mu }\left[ g_{\alpha \beta }\right] \rangle _{{\rm s s}}
\label{emtcurved2}
\end{equation}
Here the first term on the right is the vacuum energy--momentum
tensor for Rindler space in coordinates $(\tau ,r, \theta ,\phi )
$, and the second one is the situation without boundaries (pure
gravitational part). The contribution from pure gravitational part
is independent of coordinates, and so will be the same as the
stady-state case(ss) \cite{davies}.

The second term in Eq.(\ref{emtcurved2}) can be rewritten as
following
\begin{equation}
\langle T_{\nu }^{\mu }\left[ g_{\alpha \beta }\right] \rangle _{{\rm ss}%
}=-\frac{1}{2880}[\frac{1}{6} \tilde H^{(1)\mu}_{\nu}-\tilde
H^{(3)\mu}_{\nu}]
\end{equation}
 The
functions $H^{(1,3)\mu}_{\nu}$ are some combinations of curvature
tensor components (see \cite{davies}). For massless scalar field
in de Sitter space (steady state), the term is given by
\cite{{davies},{Dowk}}
\begin{equation}\label{purds}
-\frac{1}{2880}[\frac{1}{6} \tilde H^{(1)\mu}_{\nu}-\tilde
H^{(3)\mu}_{\nu}]
=\frac{1}{960\pi^{2}\alpha^{4}}\delta^{\mu}_{\nu}.
\end{equation}
Therefore the vacuum energy-momentum tensor (\ref{emtcurved2}) has
the form
\begin{equation}
\langle T_{\nu }^{\mu }\left[ g_{\alpha \beta }\right] \rangle _{{\rm ren}%
}=\langle T_{\nu }^{\mu }\left[ g_{\alpha \beta }\right] \rangle _{{\rm ren}%
}^{(0)}+\langle T_{\nu }^{\mu }\left[ g_{\alpha \beta }\right] \rangle _{%
{\rm ren}}^{(b)}.  \label{emtcurved1}
\end{equation}
Where the first term on the right is the vacuum energy--momentum
tensor for the situation without boundaries (gravitational part),
and the second one is due to the presence of boundaries. By taking
into account Eqs.(\ref{subRindm01},\ref{emtcurved2}, \ref{purds})
the first term in Eq.(\ref{emtcurved1})can be rewritten as
following
\begin{equation}
\langle T_{\nu }^{\mu }\left[ g_{\alpha \beta }\right] \rangle _{{\rm ren}%
}^{(0)}=\frac{-1}{480 \pi^{2}}(\alpha^{2}-r^{2})^{-2}
diag(1,-1/3,-1/3,-1/3)+\frac{1}{960\pi^{2}\alpha^{4}}\delta^{\mu}_{\nu}.
  \label{emtcurved11}
\end{equation}
The boundary part in Eq.(\ref{emtcurved1}) is related to the
corresponding Rindler spacetime counterpart Eq.(\ref{tmiuniu}) by
the relation
\begin{equation}
\langle T_{\nu }^{\mu }\left[ g_{\alpha \beta }\right] \rangle _{%
{\rm ren}}^{(b)}=\xi^{4}( \alpha^{2}-r^{2})^{-2}(\langle
T_{i}^{k}\rangle ^{(1b)}_D(\xi _{1},\xi )+\langle T_{i}^{k}\rangle
^{(1b)}_D(\xi _{2},\xi )+\Delta \langle T_{i}^{k}\rangle _D(\xi
_{1},\xi _{2},\xi )),\quad \xi _{1}<\xi <\xi _{2}.
\label{emtcurved12}
\end{equation}
Now we turn to the interaction forces between the plates. The
vacuum force acting per unit surface of the plate at $\xi =\xi
_{i}$ is determined by the ${}^{1}_{1}$--component of the vacuum
EMT at this point. The gravitational part of the pressure
according to Eq.(\ref{emtcurved11}) is equal to
\begin{equation}\label{grper}
P_{g}=-<T^{1}_{1}>=\frac{-1}{960\pi^{2}\alpha^{4}}-\frac{1}{1440
\pi^{2}}(\alpha^{2}-r^{2})^{-2}.
\end{equation}
The first term is the same from both sides of the plates, and
hence leads to zero effective force.  The corresponding effective
boundary part pressures can be presented as a sum of two terms (at
first we consider boundary part pressures in Rindler spacetime)
\begin{equation}
p_{b}^{(i)}=p_{b1}^{(i)}+p_{b{\rm (int)}}^{(i)},\quad i=1,2.
\label{FintD}
\end{equation}
The first term on the right is the pressure for a single plate at $%
\xi =\xi _{i}$ when the second plate is absent. This term is
divergent due to the well known surface divergences in the
subtracted VEV's. The second term on the right of Eq.
(\ref{FintD}),
\begin{equation}
p_{b{\rm (int)}}^{(i)}=-\langle T_{1}^{1}\rangle ^{(1b)}_b(\xi
_{j},\xi _{i})-\Delta \langle T_{1}^{1}\rangle _b(\xi _{1},\xi
_{2},\xi _{i}),\quad i,j=1,2,\quad j\neq i \label{pintD}
\end{equation}
is the pressure induced by the presence of the second plate, and
can be termed as an interaction force. For the plate at $\xi =\xi
_{2}$ the interaction term is due to the second summand on the
right of Eq. (\ref{EMTDdiag1}). Substituting into this term $\xi
=\xi _{2}$ and using the Wronskian relation for the modified
Bessel functions one has
\begin{equation}
p_{b{\rm (int)}}^{(2)}(\xi _1,\xi _2)=-\frac{A_3}{2\xi _{2}^{2}}
\int_{0}^{\infty }dkk\int_{0}^{\infty }d\omega \frac{%
I_{\omega }(k\xi _{1})}{I_{\omega }(k\xi _{2})D_{\omega }(k\xi
_{1},k\xi _{2})}.  \label{pint2}
\end{equation}
By a similar way  for the interaction term on the plate at $\xi
=\xi _{1}$ we obtain
\begin{equation}
p_{b{\rm (int)}}^{(1)}(\xi _1,\xi _2)=-\frac{A_3}{2\xi _{1}^{2}}
\int_{0}^{\infty }dkk\int_{0}^{\infty }d\omega \frac{%
K_{\omega }(k\xi _{2})}{K_{\omega }(k\xi _{1})D_{\omega }(k\xi
_{1},k\xi _{2})}.  \label{pint1}
\end{equation}

In the limit $\xi _2\gg \xi _1$, introducing in Eq. (\ref{pint2})
a new integration variable $x=k\xi _2$, and making use the formula
\begin{equation}\label{Ismall}
  I_\omega (y)=\left( \frac{y}{2}\right) ^{\omega }\frac{1}{\Gamma (\omega )}
  \left[ 1+{\cal O}(y^2)\right] ,\quad y=x\xi _1/\xi _2,
\end{equation}
and the standard relation between the functions $K_\omega $ and
$I_{\pm \omega }$ one finds
\begin{equation}
p_{b{\rm (int)}}^{(2)} \approx -\frac{\pi ^2A_3}{48 \xi
_{2}^{4}\ln ^2(2\xi _{2}/\xi _{1})}\int_{0}^{\infty
}\frac{dxx}{I_{0}^{2}(x)}\left[1+{\cal O}\left( \frac{\ln x }{\ln
(2\xi _2/\xi _1)} \right) \right] . \label{pD2far}
\end{equation}
The similar calculation for Eq. (\ref{pint1}) yields
\begin{equation}
p_{b{\rm (int)}}^{(1)} \approx -\frac{\pi ^2A_3}{24 \xi _{2}\xi
^2_1\ln ^3(2\xi _{2}/\xi _{1})}\int_{0}^{\infty }\frac{dxx
K_0(x)}{I_{0}(x)}\left[1+{\cal O}\left( \frac{\ln x }{\ln (2\xi
_2/\xi _1)} \right) \right] . \label{pD1far}
\end{equation}
 Therefore the effective boundary part pressures in static de
 Sitter space acting on the plates are given by, $%
p_{b(int)}^{1,2}=-\langle T_{1}^{1}\left[ g_{\alpha \beta }\right] \rangle _{{\rm ren}%
}^{(b)}$, for the plate at $\xi =\xi _{2}$ we have

\begin{equation}
p_{b{\rm (int)}}^{(2)}(\xi _1,\xi _2)=-\xi_{2}^{2}(
\alpha^{2}-r^{2})^{-2}\frac{A_3}{2}
\int_{0}^{\infty }dkk\int_{0}^{\infty }d\omega \frac{%
I_{\omega }(k\xi _{1})}{I_{\omega }(k\xi _{2})D_{\omega }(k\xi
_{1},k\xi _{2})}.  \label{pint21}
\end{equation}
For the interaction term on the plate at $\xi =\xi _{1}$ we obtain
\begin{equation}
p_{b{\rm (int)}}^{(1)}(\xi _1,\xi _2)=-\xi_{1}^{2}(
\alpha^{2}-r^{2})^{-2}\frac{A_3}{2}
\int_{0}^{\infty }dkk\int_{0}^{\infty }d\omega \frac{%
K_{\omega }(k\xi _{2})}{K_{\omega }(k\xi _{1})D_{\omega }(k\xi
_{1},k\xi _{2})}.  \label{pint13}
\end{equation}
As the function $D_{\omega }(k\xi ,k\xi _2)$ is positive for $\xi
_1<\xi _2$, interaction forces per unit surface Eqs.(\ref{pint21})
and (\ref{pint13}) are always attractive.

 \section{Conclusion}
  Quantum field theory in accelerated
systems contains many of special features produced by a
gravitational field avoiding some of the difficulties entailed by
renormalization in a curved spacetime. In particular, the near
horizon geometry of most black holes is well approximated by
Rindler and a better understanding of physical effects in this
background could serve as a handle to deal with more complicated
geometries like Schwarzschild. The Rindler geometry shares most of
the qualitative features of black holes and is simple enough to
allow detailed analysis. Another motivation for the investigation
of quantum effects in the Rindler space is related to the fact
that this space is conformally related to the de Sitter space and
to the Robertson--Walker space with negative spatial curvature. As
a result the expectation values of the energy--momentum tensor for
a conformally invariant field and for corresponding conformally
transformed boundaries on the de Sitter and Robertson--Walker
backgrounds can be derived from the corresponding Rindler
counterpart by the standard transformation .\\
 In the present
paper we have investigated the Casimir effect for a conformally
coupled massless scalar field between two parallel plates moving
by uniform acceleration, on background of the static de Sitter
spacetimes which is conformally related to the Rindler spacetime.
We have assumed that the scalar field satisfies Dirichlet boundary
condition on the plates. The vacuum expectation values of the
energy-momentum tensor are derived from the corresponding Rindler
spacetime results by using the conformal properties of the
problem. As the boundaries are static in the Rindler coordinates
no Rindler quanta are created. In the region between the plates
the boundary induced part for the vacuum energy-momentum tensor is
given by Eq.(\ref{emtcurved12}), and the corresponding vacuum
forces acting per unit surface of the plates have the form Eqs.
(\ref{pint21}),(\ref{pint13}). The vacuum polarization due to the
gravitational field, without any boundary conditions is given by
Eq.(\ref{emtcurved11}), the corresponding gravitational pressure
part has the form Eq.(\ref{grper}),the first term in this equation
is the same from both sides of the plates, and hence leads to zero
effective force. Therefore the effective force acting on the
plates are given only by the boundary part of the vacuum
pressures.\\
Our calculations may be of interest in the brane-world
cosmological scenarios. The brane-world corresponds to a manifold
with dynamical boundaries and all fields which propagate in the
bulk will give Casimir-type contributions to the vacuum energy,
and as a result to the vacuum forces acting on the branes. In
dependence of the type of a field and boundary conditions imposed,
these forces can either stabilize or destabilize the brane-world.
In addition, the Casimir energy gives a contribution to both the
brane and bulk cosmological constant and, hence, has to be taken
into account in the self-consistent formulation of the brane-world
dynamics.( see for example \cite{{set6},{set7}, {set8}, {eliz}}).

\section*{Acknowledgement }
I would like to thank Prof. A.A. Saharian for his comments and
remarks.

\end{document}